# A KIF Formalization for the IFF Category Theory Ontology

Robert E. Kent

550 Staley Dr.
Pullman, WA 99163
rekent@ontologos.org

**Abstract**

This paper begins the discussion of how the Information Flow Framework can be used to provide a principled foundation for the metalevel (or structural level) of the Standard Upper Ontology (SUO). This SUO structural level can be used as a logical framework for manipulating collections of ontologies in the object level of the SUO or other middle level or domain ontologies. From the Information Flow perspective, the SUO structural level resolves into several metalevel ontologies. This paper discusses a KIF formalization for one of those metalevel categories, the Category Theory Ontology. In particular, it discusses its category and colimit sub-namespaces.

## The Information Flow Framework

The mission of the Information Flow Framework (IFF) is to further the development of the theory of Information Flow, and to apply Information Flow to distributed logic, ontologies, and knowledge representation. IFF provides mechanisms for a principled foundation for an ontological framework – a framework for sharing ontologies, manipulating ontologies as objects, partitioning ontologies, composing ontologies, discussing ontological structure, noting dependencies between ontologies, declaring the use of other ontologies, etc.

IFF is *primarily* based upon the theory of Information Flow initiated by Barwise (Barwise and Seligman 1997), which is centered on the notion of a *classification*. Information Flow itself based upon the theory of the Chu construction of ∗-autonomous categories (Barr 1996), thus giving it a connection to concurrency and Linear Logic. IFF is *secondarily* based upon the theory of Formal Concept Analysis initiated by Wille (Ganter & Wille 1999) , which is centered on the notion of a *concept lattice*.

IFF represents metalogic, and as such operates at the structural level of ontologies. In IFF there is a precise boundary between the metalevel and the object level. The structure of IFF is illustrated in Figure 1. This consists of a collection of metalevel ontologies, usually centered on a category-theory *category* of IFF.

○ At the upper metalevel is the Basic KIF Ontology, whose purpose is to provide an interface between KIF and ontological structure. The Basic KIF Ontology provides an adequate foundation for representing ontologies in general and for defining the other metalevel ontologies of Figure 1 in particular. All ontologies import and use the Basic KIF ontology.

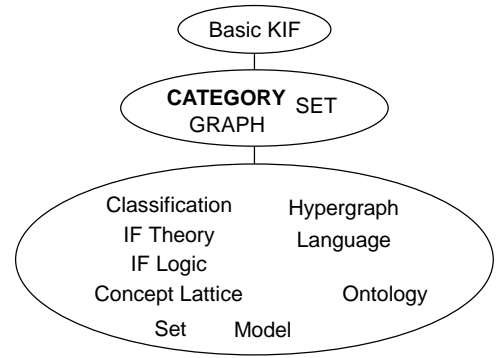

**Figure 1: Metalevel Ontologies of IFF**

○ At the middle metalevel are three generic ontologies – a *Category Theory Ontology* (partially presented in this paper) that allows us to make claims about lower metalevel ontologies such as "the Classification Ontology represents a category" or "the classification functor is left adjoint to the IF theory functor" (Kent 2000), a *GRAPH Ontology* that provides the mathematical context in which the Category Theory Ontology can be defined (as a monoid in a monoidal category), and a *SET Ontology* which provides foundational axiomatics and terminology for ontologies in the middle or lower metalevel, such as the Category Theory and Classification ontologies. The Category Theory Ontology is a KIF formalism for *category theory* in one of its normal presentations. Other presentations, such as home-set or arrows-only, may also have value. The Category Theory Ontology provides a framework for reasoning about metalogic, as represented in the lower metalevel ontologies.

○ At the lower metalevel, ontologies are organized in two dimensions, an instantiation-predication dimension and an entity-relation dimension. These two precise mathematical dimensions correspond to the intuitive distinctions of Heraclitus and Peirce.

  □ In IFF the type-token distinction looms large. In the instantiation-predication dimension are the *Classification Ontology* that directly represents the type-token distinction, the *IF Theory Ontol-* *ogy* that is connected by an adjunction to the Classification Ontology, and the combining *IF Logic Ontology*. The Classification Ontology declares and axiomatizes the central '`Classification`' construction. The IF Theory Ontology, which is based upon a sequent calculus, declares and axiomatizes the standard predicates of '`subclass`', '`disjoint`' and '`partition`'. Representing Formal Concept Analysis is a *Concept Lattice Ontology*. In addition to formal concepts and their lattices, this also includes the idea of a collective concept.

☐ In the entity-relation dimension are the *Hypergraph Ontology* that represents multivalent relations and the *Language Ontology* (whose presentation is a little delicate) that represents logical expressions.

☐ Also at the lower level are the *Set Ontology* that models the topos of small sets and is regarded to be part of the IFF foundations), the combining *Model Ontology* and a derivative *Ontology Ontology*. The latter two ontologies are related to the fundamental truth meta-classification[1] between models and expressions. In addition, the lower metalevel ontologies have sufficient morphism and colimit structure to build ontologies at the object level.

## Overview of the paper

In this paper we discuss and present the Category Theory Ontology. The Category Theory Ontology is an example of an interesting mathematical ontology – an ontology that represents a mathematical context, such as Groups, Matrices, Topological Spaces, etc. The Category Theory Ontology is a basic component in IFF that allows reasoning about both meta and object levels.

In general, each particular IFF ontology defines a namespace with (normally) sub-namespaces for the morphism and colimit aspects of the category being represented. References outside the ambient namespace can use namespace prefixing to disambiguate when necessary. For example, when referring to the category composition operation from the functor namespace, the notation '`CAT$composition`' can be used. However, for the Category Theory Ontology there are multiple namespaces, one for each of the major components of category theory: category, functor, natural transformation and adjunction.

As listed in Table 1, the Category Theory Ontology imports and uses terms from the Basic Ontology, the foundational SET Ontology and the GRAPH Ontology.

| KIF | '`function`', '`signature`', '`class`', '`subclass`' |
|---|---|
| SET | '`conglomerate`', '`class`', '`subclass`' '`function`', '`source`', '`target`' '`composition`', '`identity`', '`inclusion`' '`terminal`', '`unique`' '`opspan`', '`opvertex`', '`opfirst`', '`opsecond`' '`pullback`', '`pullback-projection1`', '`pullback-projection2`' |
| GPH | '`graph`', '`object`', '`morphism`' '`source`', '`target`', '`opposite`' '`opspan`', '`multiplication`', '`unit`' |
| GPH .MOR | '`graph-morphism`', '`source`', '`target`' '`object`', '`morphism`' '`composition`', '`identity`' '`multiplication`', '`alpha`', '`left`', '`right`', '`tau`' |

**Table 1: Terms imported into the Category Theory Ontology**

Table 2 lists the terms that are defined and axiomatized in this part of the Category Theory Ontology described in this paper. The boldface terms are used in the examples to declare and populate specific concrete categories.

| CAT | '`category`', '`underlying`', '`mu`', '`composition`', '`eta`', '`identity`' '`object`', '`morphism`', '`source`', '`target`' '`composable-opspan`', '`composable`', '`first`', '`second`', '`opposite`' '`monomorphism`', '`epimorphism`', '`isomorphism`' |
|---|---|
| COL | '`initial`', '`counique`' '`span`', '`vertex`', '`first`', '`second`' '`cocone2`', '`cocone2-opspan`' '`opvertex`', '`opfirst`', '`opsecond`' '`pushout-cocone2`', '`pushout`' '`comediator2-opspan`', '`comediator2-pair`' '`comediator2`', '`finitely-cocomplete`', '`diagram`' '`terminal`', '`unique`', '`object`', '`constant`' '`cocone`', '`colimit`', '`comediator`', '`cocomplete`' |

**Table 2: Terms originating in the Category Theory Ontology**

In addition to the IFF section and this overview section, a larger version of this paper has a section for each component of category theory – categories, functors, natural transformations, adjunctions, limits and colimits. Because of space limitations, only the category and colimit components of the Category Theory Ontology are discussed here. Although the other components are not included, some of their classes and functions are visible in the section on colimits.

## Categories

A *category* can be thought of as a special kind of graph – a graph with monoidal properties. More precisely, a cate-gory $C = \langle C, \mu_C, \eta_C \rangle$ is a monoid in the 2-dimensional quasi-category of (large) graphs and graph morphisms. This means that it consists of a graph $C$, a *composition* graph morphism $\mu_C : C \otimes C \to C$ and an *identity* graph morphism $\eta_C : 1_{obj(C)} \to C$, both with the identity object function $id_{obj(C)}$. Table 3 gives the notation for the composition function $\circ^C = mor(\mu_C)$ and the identity function $id^C = mor(\eta_C)$ – these are the morphism functions of the composition and identity graph morphisms.

---

[1] The truth classification of a first-order language *L* is the meta-classification, whose instances are *L*-structures (models), whose types are *L*-sentences, and whose classification relation is satisfaction. In IFF the concept lattice of the truth meta-classification functions is the appropriate "lattice of ontological theories." A formal concept in this lattice has an intent that is a closed theory (set of sentences) and an extent that is the collection of all models for that theory. The theory (intent) of the join or supremum of two concepts is the closure of the intersection of the theories (conceptual intents), and the theory (intent) of the meet or infimum of two concepts is the theory of the common models.

| $\circ^C : mor(C) \times_{obj(C)} mor(C) \to mor(C)$ | $id^C : obj(C) \to mor(C)$ |
|---|---|
| $(m_1, m_2) \mapsto m_1 \circ m_2$ | $o \mapsto id_o$ |

**Table 3: Elements of Monoidal Structure**

Axioms (1–4) give the KIF representation for a category. The unary KIF function 'underlying' of axiom (2) gives the *underlying* graph of a category. The SET function '(composition ?c)' of axiom (3.1) provides for an associative composition of morphisms in the category – it operates on any two morphisms that are composable, in the sense that the target object of the first is equal to the source object of the second, and returns a well-defined (composition) morphism. The SET function '(identity ?c)' in axiom (4.1) provides identities – it associates a well-defined (identity) morphism with each object in the category. The unary KIF functions 'composition' and 'identity' have the category as a parameter.

```
(1) (SET$conglomerate category)

(2) (KIF$function underlying)
    (KIF$signature underlying
     category GPH$graph)

(3) (KIF$function mu)
    (KIF$signature mu
     category GPH.MOR$graph-morphism)
    (forall (?c (category ?c))
    (and
      (= (GPH.MOR$source (mu ?c))
         (GPH$multiplication
            (underlying ?c)
            (underlying ?c)))
      (= (GPH.MOR$target (mu ?c))
         (underlying ?c))
      (= (GPH.MOR$object (mu ?c))
         (SET$identity
            (GPH$object
               (underlying ?c))))))

(3.1) (KIF$function composition)
    (KIF$signature composition
     category SET$function)
    (forall (?c (category ?c))
    (= (composition ?c)
       (GPH.MOR$morphism (mu ?c))))

(4) (KIF$function eta)
    (KIF$signature eta
     category GPH.MOR$graph-morphism)
    (forall (?c (category ?c))
    (and
      (= (GPH.MOR$source (eta ?c))
```
```
         (GPH$unit
            (GPH$object (underlying ?c))))
      (= (GPH.MOR$target (eta ?c))
         (underlying ?c))
      (= (GPH.MOR$object (eta ?c))
         (SET$identity
            (GPH$object
               (underlying ?c))))))

(4.1) (KIF$function identity)
    (KIF$signature identity
     category SET$function)
    (forall (?c (category ?c))
    (= (identity ?c)
       (GPH.MOR$morphism (eta ?c))))
```

For convenience in the language used for categories, in axioms (5–11) we rename the object and morphism classes, the source and target functions, the class of composable pairs of morphisms, and the first and second functions in the setting of categories.

```
(5) (KIF$function object)
    (KIF$signature object
     category SET$class)
    (forall (?c (category ?c))
     (= (object ?c)
        (GPH$object (underlying ?c))))

(6) (KIF$function morphism)
    (KIF$signature morphism
     category SET$class)
    (forall (?c (category ?c))
     (= (morphism ?c)
        (GPH$morphism (underlying ?c))))

(7) (KIF$function source)
    (KIF$signature source
     category SET$function)
    (forall (?c (category ?c))
     (= (source ?c)
        (GPH$source (underlying ?c))))

(8) (KIF$function target)
    (KIF$signature source
     category SET$function)
    (forall (?c (category ?c))
     (= (target ?c)
        (GPH$target (underlying ?c))))

(9) (KIF$function composable-opspan)
    (KIF$signature composable-opspan
     category SET$opspan)
    (forall (?c (category ?c))
     (= (composable-opspan ?c)
        (GPH$opspan
           (underlying ?c)
           (underlying ?c))))

(10) (KIF$function composable)
    (KIF$signature composable
     category SET$class)
    (forall (?c (category ?c))
     (= (composable ?c)
        (GPH$morphism
           (GPH$multiplication
              (underlying ?c)
              (underlying ?c)))))

(11.1) (KIF$function first)
```

```
     (KIF$signature first
      category SET$function)
     (forall (?c (category ?c))
       (= (first ?c)
          (SET$pullback-projection1
            (GPH$opspan
              (underlying ?c)
              (underlying ?c))))))
(11.2) (KIF$function second)
       (KIF$signature second
        category SET$function)
       (forall (?c (category ?c))
         (= (second ?c)
            (SET$pullback-projection2
              (GPH$opspan
                (underlying ?c)
                (underlying ?c))))))
```

By the definitions of graph morphisms, graph multiplication and graph units, these operations satisfy the typing constraints of Table 4 for any category *C*.

| |
|---|
| $\circ^C \cdot src(C) = 1^{st}(C) \cdot src(C)$ |
| $\circ^C \cdot tgt(C) = 2^{nd}(C) \cdot tgt(C)$ |
| $id^C \cdot src(C) = id_{obj(C)} = id^C \cdot tgt(C)$ |

**Table 4: Preservation of Source and Target**

Table 5 contains commutative diagrams of graph morphisms. The commutative diagram on the left in Table 3 represents the *associative law* for composition, and the commutative diagrams on the right in Table 3 represent the left and right *unit laws* for identity.

| | |
|---|---|
| $C \otimes (C \otimes C) \xrightarrow{C \otimes \mu_C} C \otimes C$ <br> $\alpha_{C,C,C} \downarrow \quad\quad\quad\quad \downarrow \mu_C$ <br> $(C \otimes C) \otimes C$ <br> $\mu_C \otimes C \downarrow$ <br> $C \otimes C \xrightarrow{\mu_C} C$ | $\eta_C \otimes C \quad\quad C \otimes \eta_C$ <br> $1 \otimes C \rightarrow C \otimes C \leftarrow C \otimes 1$ <br> $\lambda_C \searrow \downarrow \mu_C \swarrow \rho_C$ <br> $C$ |
| **Associative Law** | **Left/Right Unit Laws** |

**Table 5: Laws of Monoidal Structure**

Table 6 is derivative – it represents these laws in terms of the composition and identity functions.

| Associative law : | $(m_1 \circ^C m_2) \circ^C m_3 = m_1 \circ^C (m_2 \circ^C m_3)$ |
|---|---|
| Identity laws : | $id^C_a \circ^C m = m = m \circ^C id^C_b$ |

**Table 6: Laws of Monoidal Structure Redux**

Axiom (12) represents the associative law in KIF. This is an important axiom, since the correct expression motivated the ontology for graphs and graph morphisms, the representation of categories as monoids in the 2-dimensional category of large graphs, and in particular the coherence axiomatization. Axiom (13) represents the unit laws in KIF. Both are expressed at the level of graph morphisms. Using composition and identity, these could also be expressed at the level of SET functions.

```
(12) (forall (?c (category ?c))
       (= (GPH.MOR$composition
            (GPH.MOR$multiplication
              (GPH.MOR$identity
                (underlying ?c))
              (mu ?c))
            (mu ?c))
          (GPH.MOR$composition
            (GPH.MOR$composition
              (GPH.MOR$alpha
                (underlying ?c)
                (underlying ?c)
                (underlying ?c))
              (GPH.MOR$multiplication
                (mu ?c)
                (GPH.MOR$identity
                  (underlying ?c))))
            (mu ?c))))
(13) (forall (?c (category ?c))
       (and
         (= (GPH.MOR$composition
              (GPH.MOR$multiplication
                (eta ?c)
                (GPH.MOR$identity
                  (underlying ?c)))
              (mu ?c))
            (GPH.MOR$left
              (underlying ?c)))
         (= (GPH.MOR$composition
              (GPH.MOR$multiplication
                (GPH.MOR$identity
                  (underlying ?c))
                (eta ?c))
              (mu ?c))
            (GPH.MOR$right
              (underlying ?c)))))
```

**Additional Categorical Structure**

Particular categories may have additional structure. This is true for the categories expressed by the IFF lower metalevel ontologies (Figure 1). Here is the KIF formalization for some of this additional structure.

To each category *C*, there is an *opposite category* $C^{op} = \langle C, \mu_C, \eta_C \rangle^{op} = \langle C^{op}, \tau_{C,C} \cdot \mu_C^{op}, \eta_C^{op} \rangle$. Since all categorical notions have their duals, the opposite category can be used to decrease the size of the axiom set. The objects of $C^{op}$ are the objects of *C*, and the morphisms of $C^{op}$ are the morphisms of *C*. However, the source and target of a morphism are reversed: $src(C^{op})(m) = tgt(C)(m)$ and $tgt(C^{op})(m) = src(C)(m)$. The composition is defined by $m_2 \circ^{op} m_1 = m_1 \circ m_2$, and the identity is $id^{op}_o = id_o$. The type restriction axioms in (14) specify the opposite operation on categories.

```
(14) (KIF$function opposite)
     (KIF$signature opposite category category)
     (forall (?c (category ?c))
       (and
         (= (underlying (opposite ?c))
```

```
         (GPH$opposite (underlying ?c))
     (= (mu (opposite ?c))
        (GPH.MOR$composition
           (GPH.MOR$tau
              (underlying ?c)
              (underlying ?c))
           (GPH.MOR$opposite (mu ?c))))
     (= (eta (opposite ?c))
        (GPH.MOR$opposite (eta ?c)))))
```

Part of the fact that opposite forms an involution is the theorem (15) that $(C^{op})^{op} = C$.

```
(15) (forall (?c (category ?c))
         (= (opposite (opposite ?c)) ?c))
```

A morphism $m : o_0 \rightarrow o_1$ is a *monomorphism* (Axiom 16) in a category $C$ when it is right-cancellable – for any two parallel morphisms $m_0, m_1 : o_2 \rightarrow o_0$, the equality $m_0 \circ^C m = m_1 \circ^C m$ implies $m_0 = m_1$. There are both abstract and concrete methods for specifying the notions of monomorphism and epimorphism. The latter requires the axiom for specific (concrete) Cartesian pullbacks in foundations. Dually, a morphism $m : o_0 \rightarrow o_1$ is an *epimorphism* in a category $C$ when it is left-cancellable – that is, when it is a monomorphism in $C^{op}$. Axiom (17) uses the duality of the opposite category to express epimorphisms. A morphism is an *isomorphism* (Axiom 18) in a category $C$ when it is both a monomorphism and an epimorphism.

```
(16) (KIF$function monomorphism)
     (KIF$signature monomorphism
       category SET$class)
     (forall (?c (category ?c))
        (SET$subclass
           (monomorphism ?c)
           (morphism ?c)))
     (forall (?c (category ?c) ?m)
     (<=>
        ((monomorphism ?c) ?m)
        (and
           ((morphism ?c) ?m)
           (forall (?m0 ?m1
              ((composable ?c) [?m0 m])
              ((composable ?c) [?m1 m]))
           (=>
              (= ((composition c?) [?m0 m])
                 ((composition c?) [?m1 m]))
              (= ?m0 ?m1))))))

(17) (KIF$function epimorphism)
     (KIF$signature epimorphism
       category SET$class)
     (forall (?c (category ?c)
        ((morphism ?c) ?m))
     (<=>
        ((epimorphism ?c) ?m)
        ((monomorphism (opposite ?c)) ?m)))

(18) (KIF$function isomorphism)
     (KIF$signature isomorphism
       category SET$class)
     (forall (?c (category ?c))
     (<=>
        ((isomorphism ?c) ?m)
        (and
           ((monomorphism ?c) ?m)
           ((epimorphism ?c) ?m))))
```

### Functors
⟨snip⟩

### Natural Transformations
⟨snip⟩

### Adjunctions
⟨snip⟩

### Colimits

Colimits are important for manipulating and composing ontologies expressed in the object language. The use of colimits advocates a "building blocks approach" to ontology construction. Continuing this metaphor, the colimits approach understands that the mortar between the ontological blocks must be strong and resilient in order to adequately support the ontological building, and requests that methods for composing component ontologies, such as merging, mapping and aligning ontologies, be made very explicit so that they can be analyzed. A compact but detailed discussion of Information Flow, with applications to this building blocks approach to ontology construction, is given in the paper (Kent 2000).

The colimit part of the Category Theory Ontology is a separate namespace. To completely express colimits, we need elements from all the basic components of category theory – categories, functors, natural transformations and adjunctions. As such, colimits provide a glimpse of other the parts of the Category Theory Ontology not contained in this paper because of space limitations. These basic components are indicated by the namespace prefixes in the KIF formalism below.

**Finite colimits in a category**

Here we present axioms for the finite cocompleteness of categories.

An *initial object* in a category is an object from which there is exactly one morphism to any other object.

```
(19) (KIF$function initial)
     (KIF$signature initial
       CAT$category SET$class)

     (forall (?c (CAT$category ?c))
        (SET$subclass
           (initial ?c)
           (CAT$object ?c)))

(19.c) (KIF$function counique)
       (KIF$signature counique
         CAT$category SET$function)

     (forall (?c (CAT$category ?c))
     (and
        (= (SET$source (counique ?c))
           (IFF$product2
              (initial ?c)
              (CAT$object ?c)))
        (= (SET$target (counique ?c))
```

```
       (CAT$morphism ?c))))

   (forall (?c (CAT$category ?c)
            ?i ((initial ?c) ?i)
            ?o ((CAT$object ?c) ?o))
     (= ((counique ?c) [?i ?o])
        (the (?m ((CAT$morphism ?c) ?m))
         (and
           (= ((CAT$source ?c) ?m) ?i)
           (= ((CAT$target ?c) ?m) ?o)))))
```

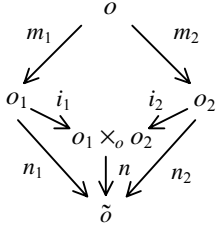

**Figure 2: Pushout**

Each span in a category consists of a pair of morphisms called *first* and *second*. These are required to have a common source object, denoted as the *vertex*. Let '`(span ?c)`' be the KIF term that denotes the *Span* class in a category.

```
(20) (KIF$function span)
     (KIF$signature span
      CAT$category IFF$class)

(21) (KIF$function vertex)
     (KIF$signature vertex
      CAT$category IFF$function)
     (forall (?c (CAT$category ?c))
      (and
        (= (IFF$source (vertex ?c))
           (span ?c))
        (= (IFF$target (vertex ?c))
           (object ?c))))

(21.1) (KIF$function first)
       (KIF$signature first
        CAT$category IFF$function)
       (forall (?c (CAT$category ?c))
        (and
          (= (IFF$source (first ?c))
             (span ?c))
          (= (IFF$target (first ?c))
             (CAT$morphism ?c))
          (= (IFF$composition
               (first ?c)
               (source ?c))
             (vertex ?c))))

(21.2) (KIF$function second)
       (KIF$signature second
        CAT$category IFF$function)
       (forall (?c (CAT$category ?c))
        (and
          (= (IFF$source (second ?c))
             (span ?c))
          (= (IFF$target (second ?c))
             (CAT$morphism ?c))
          (= (IFF$composition
               (second ?c)
               (source ?c))
             (vertex ?c))))
```

*Binary cocones* in categories are used to specify and axiomatize finite colimits. Each binary cocone has an underlying *span*, a *opvertex* object, and a pair of morphisms called *opfirst* and *opsecond*, whose common target object in the category is the opvertex and whose source objects are the target objects of the morphisms in the span. The opfirst and opsecond morphisms form a commutative diagram with the span. This is a very special case of a cocone under a diagram. Let '`(cocone2 ?c)`' be the KIF term that denotes the *Binary Cocone* class in the category.

```
(22) (KIF$function cocone2)
     (KIF$signature cocone2
      CAT$category IFF$class)

(23) (KIF$function cocone2-span)
     (KIF$signature cocone2-span
      CAT$category IFF$function)
     (forall (?c (CAT$category ?c))
      (and
        (= (IFF$source (cocone2-span ?c))
           (cocone2 ?c))
        (= (IFF$target (cocone2-span ?c))
           (span ?c))))

(24) (KIF$function opvertex)
     (KIF$signature opvertex
      CAT$category IFF$function)
     (forall (?c (CAT$category ?c))
      (and
        (= (IFF$source (opvertex ?c))
           (cocone2 ?c))
        (= (IFF$target (opvertex ?c))
           (object ?c))))

(24.1) (KIF$function opfirst)
       (KIF$signature opfirst
        CAT$category IFF$function)
       (forall (?c (CAT$category ?c))
        (and
          (= (IFF$source (opfirst ?c))
             (cocone2 ?c))
          (= (IFF$target (opfirst ?c))
             (CAT$morphism ?c))
          (= (IFF$composition
               (opfirst ?c)
               (target ?c))
             (opvertex ?c))))

(24.2) (KIF$function opsecond)
       (KIF$signature opsecond
        CAT$category IFF$function)
       (forall (?c (CAT$category ?c))
        (and
          (= (IFF$source (opsecond ?c))
             (cocone2 ?c))
          (= (IFF$target (opsecond ?c))
             (CAT$morphism ?c))
          (= (IFF$composition
               (opsecond ?c)
               (target ?c))
             (opvertex ?c))))

(25) (forall (?c (CAT$category ?c))
      (= (IFF$composition
          (IFF$pairing (composable-opspan ?c)
            (IFF$composition
```

```
            (cocone2-span ?c)
              (first ?c))
           (opfirst ?c))
          (CAT$composition ?c))
         (IFF$composition
          (IFF$pairing (composable-opspan ?c)
           (IFF$composition
            (cocone2-span ?c)
              (second ?c))
            (opsecond ?c))
          (CAT$composition ?c))))
```

*Pushouts cocones* (Figure 2) in categories are special binary cocones. They are universal in the sense that from the opvertex of any other binary cocone that shares a common span with a pushout, there is a unique morphism called the mediator morphism, that commutes with opfirst and opsecond morphism.

The opvertex object of the pushout cocone is given by the KIF function 'pushout'. It comes equipped with two KIF injection morphisms 'pushout-injection1' and 'pushout-injection2'. This notation is for convenience of reference. It is used for pushouts in categories in general.

```
(26) (KIF$function pushout-cocone2)
     (KIF$signature pushout-cocone2
      CAT$category IFF$class)
     (forall (?c (CAT$category ?c))
      (SET$subclass
        (pushout-cocone2 ?c)
        (cocone2 ?c)))

     (forall (?c (CAT$category ?c)
              ?p ((pushout-cocone2 ?c) ?p)
              ?s ((cocone2 ?c) ?s))
      (=>
       (= ((cocone2-span ?c) ?s)
          ((cocone2-span ?c) ?p))
       (exists-unique (?m
        ((CAT$morphism ?c) ?m))
       (and
        ((CAT$composable ?c)
          [((opfirst ?c) ?p) ?m])
        (= ((CAT$composition ?c)
            [((opfirst ?c) ?p) ?m])
           ((opfirst ?c) ?s))
        ((CAT$composable ?c)
         [((opsecond ?c) ?p) ?m])
        (= ((CAT$composition ?c)
            [((opsecond ?c) ?p) ?m])
           ((opsecond ?c) ?s))))))

(26.p) (KIF$function pushout)
       (KIF$signature pushout
        CAT$category SET$class)
       (forall (?c (CAT$category ?c))
         (SET$subclass
          (pushout ?c)
          (object ?c)))

       (forall (?c (category ?c))
        (<=>
         ((pushout ?c) ?o)
         (exists (?p ((pushout-cocone2 ?c) ?p)
          (= ?o ((opvertex ?c) ?p)))))

(26.o) (KIF$function comediator2-opspan)
       (KIF$signature comediator2-opspan
        CAT$category SET$opspan)

       (forall (?c (CAT$category ?c))
        (and
         (= (SET$opvertex (comediator2-opspan ?c))
            (span ?c))
         (= (SET$opfirst (comediator2-opspan ?c))
            (SET$composition
             (SET$inclusion
              (pushout-cocone2 ?c)
              (cocone2 ?c))
             (cocone2-span ?c)))
         (= (SET$opsecond (comediator2-opspan ?c))
            (cocone2-span ?c))))

(26.pr) (KIF$function comediator2-pair)
        (KIF$signature comediator2-pair
         CAT$category SET$class)

        (forall (?c (CAT$category ?c))
         (= (comediator2-pair ?c)
            (SET$pullback
             (comediator2-opspan ?c))))

(26.c) (KIF$function comediator2)
       (KIF$signature comediator2
        CAT$category SET$function)

       (forall (?c (CAT$category ?c))
        (and
         (= (SET$source (comediator2 ?c))
            (comediator2-pair ?c))
         (= (SET$target (comediator2 ?c))
            (CAT$morphism ?c))))

       (forall (?c (CAT$category ?c)
         ?p ?s ((comediator2-pair ?c) [?p ?s]))
         (= ((comediator2 ?c) [?p ?s])
            (the (?m ((CAT$morphism ?c) ?m))
             (and
              ((CAT$composable ?c)
               [((opfirst ?c) ?p) ?m])
              (= ((CAT$composition ?c)
                  [((opfirst ?c) ?p) ?m])
                 ((opfirst ?c) ?s))
              ((CAT$composable ?c)
               [((opsecond ?c) ?p) ?m])
              (= ((CAT$composition ?c)
                  [((opsecond ?c) ?p) ?m])
                 ((opsecond ?c) ?s))))))
```

A category is finite cocomplete[2] when it has an initial object, it has binary coproducts, it has a coequalizer for any parallel pair of morphisms, and it has a pushout for any binary cocone. These notions are not independent – pushouts can be defined in terms of binary coproducts and coequalizers, and binary coproducts can be defined in terms of pushouts and initiality.

```
(27) (KIF$class finitely-cocomplete)
     (KIF$subclass
        finitely-cocomplete
        CAT$category)

     (forall (?c (CAT$category ?c))
```

---

[2] Because of space limitations two other colimit notions are not defined in this paper: *binary coproduct* and *coequalizer*.

```
    (<=>
      (finitely-cocomplete ?c)
      (and
        (not (empty (initial ?c)))
        (forall (?pp ((parallel-pair ?c) ?pp))
          (exists (?m
            ((coequalizer ?c) ?m))
            (= ((source ?c) ?m)
               ((target ?c) ?pp))))
        (forall (?s ((cocone2 ?c) ?s))
          (exists (?p
            ((pushout-cocone2 ?c) ?p))
            (= ((cocone2-span ?c) ?p)
               ((cocone2-span ?c) ?s))))))))
```

**General colimits in a category**

Given a category *C* a *diagram D* in the category *C* is a functor from some "shape" category *J* to *C*. Axiom (28) defines a diagram in a category. The *shape* of a diagram is its source category.

```
(28) (KIF$function diagram)
     (KIF$signature diagram
      CAT$category KIF$class)
     (forall (?c (CAT$category ?c))
       (KIF$subclass (diagram ?c) FUNC$functor))
     (forall (?c (CAT$category ?c)
              ?d ((diagram ?c) ?d))
       (= (FUNC$target ?d) ?c))
```

The terminal category **1** defined in axiom (29) has one object and one morphism.

```
(29) (CAT$category terminal)
     (= (CAT$object terminal) SET$terminal)
     (= (CAT$morphism terminal) SET$terminal)
```

Given any category *C*, there is a unique functor $!_C : C \to \mathbf{1}$ called the *unique functor* from *C*, whose object and morphism functions are defined in axiom (30) in terms of the SET unique object and morphism functions.

```
(30) (KIF$function unique)
     (KIF$signature unique
      CAT$category FUNC$Functor)
     (forall (?c (CAT$category ?c))
     (and
       (= (FUNC$source (unique ?c)) ?c)
       (= (FUNC$target (unique ?c)) terminal)))
     (forall (?c (CAT$category ?c))
     (and
       (= (FUNC$object (unique ?c))
          (SET$unique (CAT$object ?c)))
       (= (FUNC$morphism (unique ?c))
          (SET$unique (CAT$morphism ?c)))))
```

Any object $o \in Obj(C)$ in a category *C* corresponds to an *object functor* $obj_C(o) : \mathbf{1} \to C$ that maps *0* to *o*. Axiom (31) defines this explicitly.

```
(31) (KIF$function object)
     (KIF$signature object
      CAT$category SET$function)
     (forall (?c (CAT$category ?c))
     (and
       (= (SET$source (object ?c))
          (CAT$object ?c))
       (= (SET$target (object ?c))
          FUNC$functor)))
     (forall (?c (CAT$category ?c))
              ?o ((CAT$object ?c) ?o))
     (and
       (= (FUNC$source ((object ?c) ?o))
          terminal)
       (= (FUNC$target ((object ?c) ?o)) ?c)))
     (forall (?c (CAT$category ?c))
              ?o ((CAT$object ?c) ?o))
     (and
       (= ((FUNC$object
            ((object ?c) ?o)) terminal #0) ?o)
       (= ((FUNC$morphism
            ((object ?c) ?o)) terminal#00)
          ((CAT$identity ?c) ?o))))
```

Given any category *C* and any category *J* used as a colimit shape category, an object $o \in obj(C)$ has an associated *constant functor* $\Delta_{J, C}(o) : J \to C$, which is defined as the functor composition $\Delta_{J, C}(o) = !_J \cdot obj_C(o)$ – it maps each object $j \in obj(J)$ to the object $o \in obj(C)$ and maps each morphism $n \in mor(J)$ to the identity morphism at *o*. Axiom (32) defines this as the functor composite.

```
(32) (KIF$function constant)
     (KIF$signature constant
      CAT$category CAT$category KIF$function)
     (forall (?j (CAT$category ?j)
              ?c (CAT$category ?c))
     (and
       (= (SET$source (constant ?j ?c))
          (CAT$object ?c))
       (= (SET$target (constant ?j ?c))
          FUNC$functor)))
     (forall (?j (CAT$category ?j)
              ?c (CAT$category ?c)
              ?o ((CAT$object ?c) ?o))
     (and
       (= (FUNC$source ((constant ?j ?c) ?o)) ?j)
       (= (FUNC$target ((constant ?j ?c) ?o)) ?c)
     ))
     (forall (?j (CAT$category ?j)
              ?c (CAT$category ?c)
              ?o ((CAT$object ?c) ?o))
       (= ((constant ?c ?j) ?o)
          (FUNC$composition
            (unique ?c)
            ((object ?c) ?o))))
```

Given a category *C* and a diagram *D* in the category *C* of some shape $J = src(D)$, a *cocone* $\tau : D \Rightarrow \Delta_{C, J}(o) : J \to C$ is a natural transformation from *D* to the constant functor $\Delta_{C, J}(o)$ for some *vertex* object $o \in obj(C)$.

```
(33) (KIF$function cocone)
     (KIF$signature cocone
      (CAT$category ?c) KIF$function)
     (forall (?c (CAT$category ?c))
       (KIF$signature (cocone ?c)
        (diagram ?c) (object ?c) (KIF$class)))
     (forall (?c (CAT$category ?c)
              ?d ((diagram ?c) ?d)
              ?o ((object ?c) ?o))
     (and
       (KIF$subclass
         ((cocone ?c) ?d ?o)
         NAT$natural-transformation)
       (forall ?n (((cocone ?c) ?d ?o) ?n))
```

```
    (and
    (= (NAT$source ?n) ?d)
    (= (NAT$target ?n
       ((constant (FUNC$source ?d) ?c) ?o))))
```

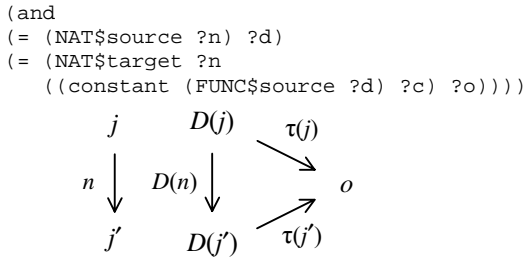

**Figure 3: Colimit Cocone**

Given a category *C* and a diagram *D* in the category *C* of some shape $J = src(D)$, a *colimit* of *D* in *C* is a universal cocone – it consists of an object $o \in obj(C)$ and a cocone
$$\gamma : D \Rightarrow \Delta_{C,J}(o) : J \to C$$
that is *universal*: for any other cocone
$$\tau : D \Rightarrow \Delta_{C,J}(o) : J \to C$$
with the same base diagram, there is a unique morphism $m : c \to o$ with $\gamma(j) \cdot m = \tau(j)$ for any indexing object $j \in Obj(J)$. Let $colim_C(D) \subseteq obj(C)$ denote the collection of all colimits of a diagram *D* in the category *C*. This may be empty. If it is nonempty for any diagram of shape *J*, then C is said to have *J-colimits*. A category C is *cocomplete*, if it has *J-colimits* for any shape *J*. For any diagram D, let $comediator_C(D)$ denote the function from $colim_C(D)$ that takes any colimit of *D* and returns its universal cocone. Axioms (34–35) formalize colimits/comediaters.

```
(34) (KIF$function colimit)
     (KIF$signature colimit
      (CAT$category KIF$function))
     (forall (?c ?d ?p (CAT$category ?c))
      (KIF$signature (colimit ?c)
       (diagram ?c) SET$class))

     (forall (?c ?d)
      (CAT$category ?c) ((diagram ?c) ?d))
      (SET$subclass
       ((colimit ?c) ?d)
       (CAT$object ?c)))

(35) (KIF$function comediator)
     (KIF$signature colimit
      (CAT$category KIF$function))
     (forall (?c (CAT$Category ?c))
      (KIF$signature (comediator ?c)
       (diagram ?c) (IFF$function)))

     (forall (?c ?d
      (CAT$category ?c)) ((diagram ?c) ?d))
     (and
      (= (IFF$source ((comediator ?c) ?d))
         ((cocone ?c) ?d))
      (= (IFF$target ((comediator ?c) ?d))
         ((colimit ?c) ?d))))

     (forall (?c ?d ?col)
      (CAT$category ?c) ((diagram ?c) ?d)
      (((colimit ?c) ?d) ?col))
       (forall (?o ?tau)
        (CAT$object ?col) ?o)
        (((cocone ?c) ?d ?o) ?tau))
         (exists-unique (?m)
```

```
          ((CAT$morphism ?c) ?m))
          (forall (?j)
           ((CAT$object (FUNC$source ?d)) ?j))
           (= ((CAT$composition ?c)
               [((NAT$component
                  (((comediator ?c) ?d) ?col)
                  ?j)
                 ?m])
              ((NAT$component ?tau) ?j))))))))
```

A category *C* is *cocomplete*, if it has colimits for any diagram of *C*. Axiom (36) formalizes cocompleteness.

```
(36) (KIF$class cocomplete)
     (KIF$subclass cocomplete CAT$category)
     (forall (?c (CAT$category ?c))
      (<=>
       (cocomplete ?c)
       (forall (?d ((diagram ?c) ?d))
        (exists (?o) (((colimit ?c) ?d) ?o))))
```

It is a standard theorem that given a category *C* and a diagram *D* in the category *C*, any two colimits are isomorphic. Axiom (37) expresses this in an external namespace.

```
(37) (forall (?c ?d ?o1 ?o2)
      (CAT$category ?c) ((COL$diagram ?c) ?d))
      (((COL$colimit ?c) ?d) ?o1)
      (((COL$colimit ?c) ?d) ?o2))
       (exists-unique (?m)
        ((CAT$isomorphism ?c) ?m)
        (and
         (= ((CAT$source ?c) ?m) ?o1)
         (= ((CAT$target ?c) ?m) ?o2))))
```

The general KIF formulation for colimits can be related to two special cases of colimits: initial objects and pushouts.

If *D* is the empty diagram in the category *C*, then a

$$0 \xrightarrow{!_o} o$$

**Figure 4: Initial object as a colimit**

colimit object, if it exists, is an initial object *0*; then, the mediator function (Figure 4) is the counique function.

The shape category **span** $= \bullet \leftarrow \bullet \rightarrow \bullet$ consists of a pair of morphisms $a_1$ and $a_2$ with common source object 0 and target objects 1 and 2, respectively. The class of objects is $\{0, 1, 2\}$, and the class of morphisms is $\{00, 11, 22, a_1, a_2\}$, with 00, 11 and 22 being the identity morphisms at objects 0, 1 and 2 respectively. The following KIF represents **span**.

```
(38) (category span)
     ((object span) span#0)
     ((object span) span#1)
     ((object span) span#2)
     ((morphism span) span#00)
     ((morphism span) span#11)
     ((morphism span) span#22)
     ((morphism span) span#a1)
     ((morphism span) span#a2)
     (= ((source span) span#a1) span#0)
     (= ((target span) span#a1) span#1)
     (= ((source span) span#a2) span#0)
     (= ((target span) span#a2) span#2)
```

```
(= ((identity span) span#0) span#00)
(= ((identity span) span#1) span#11)
(= ((identity span) span#2) span#22)
```

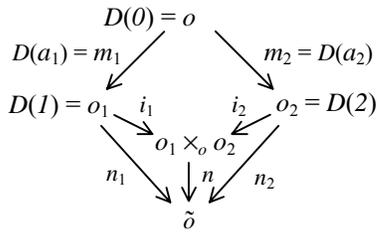

**Figure 5: Pushout as colimit**

If $D$ is a $C$-span, that is, a diagram in the category $C$ of shape **span**, whose object function maps *0*, *1* and *2* to the $C$-objects $o$, $o_1$ and $o_2$, respectively, and whose morphism function maps $a_0$ and $a_1$ to the $C$-morphisms $m_0$ and $m_1$, respectively, then a colimit object, if it exists, is a pushout $o_1 \times_o o_2$ (and vice-versa); in this case, the pushout binary cocone (Figure 5) has as its opfirst morphism the pushout injection $i_1 : o_1 \to o_1 \times_o o_2$ and has as opsecond morphism the pushout injection $i_2 : o_2 \to o_1 \times_o o_2$.

## Examples

Here we show how the Category Theory Ontology can be used as a language for representing categories in general, and lower metalevel ontologies in particular.

The IFF Classification Ontology is a lower metalevel ontology of IFF (Figure 1), which principally represents the Classification category. A preliminary KIF formalization of the Classification Ontology has been submitted to the SUO list. However, the natural language assertions that "Classification is a category" and that "Classification is cocomplete" could not be made in this preliminary version, since not enough categorical machinery was present. With the Category Theory Ontology we have that machinery. The following KIF formalization expresses both assertions in an external namespace (`cls` is a namespace prefix for the Classification Ontology in general, and `cls.info` is a namespace prefix for its morphism part).

The first group of assertions (39) centers on the claim that the term `Classification` represents a bona fide category. To make that assertion requires that we also describe or identify the components of a category: object and morphism sets, source and target functions, the class of composable pairs of infomorphisms equipped with its two projection functions, and identity and composition functions. Note that in the second statement of (39), the object set identification, there are two meanings for the string `classification`: the name of a category in an external namespace, and the name of the classification class within the Classification Ontology namespace. In order to verify these assertions, some theorem proving needs to be done. Proofs of lemmas for these assertions, such as the associativity of `cls.info$composition`, involve getting further into the details of the specific category, in this case Classification; in particular, the instance and type function components of an infomorphism, and the associativity of `set.ftn$composition`. The second group of assertions (40) claims that the category represented by `classification` is cocomplete – has all colimits. This also requires some theorem proving.

```
(39) (CAT$category classification)
     (= (CAT$object classification)
        Cls$classification)
     (= (CAT$morphism classification)
        cls.info$infomorphism)
     (= (CAT$source classification)
        cls.info$source)
     (= (CAT$target classification)
        cls.info$target)
     (= (CAT$composable classification)
        cls.info$composable)
     (= (CAT$first classification)
        cls.info$first)
     (= (CAT$second classification)
        cls.info$second)
     (= (CAT$composition classification)
        cls.info$composition)
     (= (CAT$identity classification)
        cls.info$identity)

(40) (forall ?r (cls.col$span ?r))
        ((CAT$pushout classification)
           (cls.col$pushout ?r)))
     (forall (?d)
     (<=>
       ((COL$diagram classification) ?d)
       (cls.col$diagram (graph ?d))))
     (COL$cocomplete classification)
     (forall (?d
       ((COL$diagram classification) ?d))
     (<=>
       ((COL$colimit classification) ?d)
       (cls.col$colimit (graph ?d))))
     (COL$cocomplete classification)
```